\documentclass[draft]{amsart}
\usepackage{amssymb,amscd}

\address{Sobolev Institute of Mathematics, Novosibirsk, Russia,
\quad
\emph
\newline\indent Novosibirsk State University,
\quad
\emph{and}
\newline\indent Laboratory of Geometric Methods in Mathematical Physics, Moscow State University}
\email{mironov@math.nsc.ru}

\newtheorem{theorem}{Theorem}[section]

\theoremstyle{definition}

\begin{document}

\title[Commuting higher rank ordinary differential operators]{Commuting higher rank ordinary differential operators}

\author[Andrey E. Mironov]{Andrey E. Mironov}
\thanks{The author is grateful to
the Hausdorff Research Institute for Mathematics (Bonn) for
hospitality during the writing of this paper. The work was also partially supported by a grant MD-5134.2012.1 from the
President of Russia; and by a grant from Dmitri Zimin's "Dynasty" foundation.}

\begin{abstract}
The theory of commuting ordinary differential operators was developed from the beginning of the XX century on.
The problem of finding commuting differential operators is solved for the case of operators of rank one. For
operators of rank greater than one it is still open. In this paper we discuss some results related to operators of rank greater than one.
\end{abstract}



\maketitle

\section{Introduction}
The commutativity condition $L_1L_2=L_2L_1$ of two differential
operators
$$
 L_n=\partial_x^n+\sum_{i=0}^{n-2}u_{i}(x)\partial_x^i,\
 \
 L_m=\partial_x^m+\sum_{i=0}^{m-2}v_{i}(x)\partial_x^{i}\
$$
is equivalent to a complicated system of nonlinear differential equation on coefficients $u_i(x),v_i(x)$.
The theory of commuting ordinary differential operators was first developed in the beginning of the
XX century in the works of Wallenberg \cite{W}, Schur \cite{S}, and Burchnall, Chaundy \cite{BC}.
Let me recall these classical results.

Wallenberg \cite{W} found operators
for the case of $n=2,m=3$
\begin{equation}\label{e1}
 L_1=\partial_x^2+u(x),\qquad L_2=\partial_x^3+\left(\frac{s_2}{4}+\frac{3}{2}u(x)\right)\partial_x+\frac{3}{4}u'(x),
\end{equation}
where $u(x)$ satisfies the integrable equation
$$
 (u')^2+2u^3+s_2u^2+s_1u+s_0=0,\ s_i\in{\mathbb C}.
$$

Let $L_k$ be an operator of order $k\geq 1$. Schur \cite{S} proved the following theorem.
\begin{theorem}
 If $L_nL_k=L_kL_n$ and $L_mL_k=L_kL_m$, then $L_nL_m=L_mL_k$.
\end{theorem}
To prove this theorem Schur used pseudo-differential operators.
There is a pseudo-differential operator
$
 S=1+s_{1}(x)\partial_x^{-1}+s_{2}(x)\partial_x^{-2}+\dots
$
such that $\hat{L}_k=S^{-1}L_kS=\partial_x^k.$ Pseudo-differential operator $\hat{L}_n=S^{-1}L_nS$ commutes with
$\hat{L}_k=\partial_x^k.$ This is possible only if $\hat{L}_n$ has constant coefficients. Similarly, $\hat{L}_m=S^{-1}L_mS$ has constant coefficients.
This means that
$\hat{L}_n\hat{L}_m=\hat{L}_m\hat{L}_n$ and hence $L_nL_m=L_mL_n$.

The next theorem was proved by
Burchnall and Chaundy \cite{BC}.

\begin{theorem}
 If $L_nL_m=L_mL_n$, then there is a nonzero polynomial $R(z,w)$ such that $R(L_n,L_m)=0$.
\end{theorem}
For example, for Wallenberg's operators the polynomial $R$ has the form
$$
 R(z,w)=w^2-\left(z^3+\frac{s_2}{2}z^2+\frac{s_2^2+2s_1}{16}z+\frac{s_1s_2-2s_0}{32}\right).
$$
The curve $\Gamma=\{(z,w)\in{\mathbb C}^2:R(z,w)=0\}$ is called the {\it spectral curve}. The spectral curve
parametrizes common eigenvalues of $L_n$ and $L_m$. If
$$
 L_n\psi=z\psi,\ L_m\psi=w\psi,
$$
then $(z,w)\in\Gamma$. The dimension $l$ of the space of common eigenfunctions of $L_n,L_m$ for fixed $z,w$ is called the {\it rank}. The number
$l$ is the same for general $P=(z,w)\in\Gamma$. The rank equals the greatest common divisor of $n$ and $m$.

Commutative rings of ordinary differential operators were
classified by Kriche\-ver \cite{K1}, \cite{K}. The ring is
determined by spectral data.
If the rank is one, then the spectral data define  commuting operators
by explicit formulas (see \cite{K1}). In the case of operators of rank greater than one there are the following results.
Krichever and Novikov \cite{KN1}, \cite{KN2} found operators of rank two corresponding to elliptic spectral curves.
Mokhov \cite{Mokh} found operators of rank three also corresponding
to elliptic spectral curves. There are also examples of operators of rank grater than one corresponding to spectral curves of genus $g>1$
(see \cite{M1}--\cite{Z}).

In Section 2 we recall the construction of rank one operators and give several examples.
In Section 3 we recall the method of deformations of Tyurin parameters and some results about operators of rank two corresponding to elliptic
spectral curves. Self-adjoint operators of rank two corresponding to hyperelliptic spectral curves of arbitrary genus are considered in Section 4.
Some open problems related to operators of rank greater than one are discussed in Section 5.

\section{Commuting differential operators of rank one}

Let the spectral curve $\Gamma$ be a compact Riemann surface. In the case of operators of rank one common eigenfunction
$\psi(x,P), P\in\Gamma$ of $L_n$ and $L_m$ has the following properties.

1) Function $\psi$ has one essential singularity at a fixed point $q\in\Gamma$
$$
 \psi=e^{kx}\left(1+\frac{\xi_1(x)}{k}+\frac{\xi_2(x)}{k^2}+\dots\right),
$$
where $k^{-1}$ is a local parameter in a neighbourhood of $q$.

2) Function $\psi$ has simple poles at some points $\gamma_1,\dots,\gamma_g$, where $g$ is the genus of $\Gamma$.

The set $\{\Gamma,q,k^{-1},\gamma_1,\dots,\gamma_g\}$
is called the {\it spectral data}. If we take the spectral data where $\gamma_1+\dots+\gamma_g$
is a non-special divisor, then there is a unique function $\psi(x,P)$ satisfying the conditions 1) and 2).
The function $\psi(x,P)$ is called the {\it Baker--Akhiezer function}.
Let us take a meromorphic function $f(P)$ on $\Gamma$ with the unique pole of order $n$ at $q.$ There is a
differential operator $L(f)$ of order $n$ such that
$$
 L(f)\psi(x,P)-f(P)\psi(x,P)=(\partial_x^n+u_{n-2}(x)\partial_x^{n-2}+\dots+u_0(x))\psi(x,P)=e^{xk}O(1/k).
$$
Coefficients $u_i(x)$ depend on $\xi_i(x).$ If the right hand side of the last equality is not zero, then we can add it to the Baker--Akhiezer
function and get a new function which has the same properties 1) and 2). But this is impossible, because the Baker--Akhiezer function is unique. Hence,
$
 L(f)\psi=f(P)\psi.
$
Similarly, for the meromorphic function $g(P)$ with the unique pole of order $m$ at $q$ we have $L(g)\psi=g(P)\psi$. Operators $L(f)$ and $L(g)$ commute because
the commutator has an infinite dimensional kernel
$
 [L(f),L(g)]\psi(x,P)=0.
$
So, every commutative ring of operators of rank one with the nonsingular spectral curve
corresponds to the spectral data, and the spectral data define a commutative ring of ordinary differential operators.

The Baker--Akhiezer function can be expressed via the theta-function of the Jacobi variety of $\Gamma$. Let us choose a basis of circles $a_i,b_i,1\leq i\leq g$ on
$\Gamma$ with the indices of intersections:
$$a_i\circ a_j=0,\ b_i\circ b_j=0,\ a_i\circ b_j=\delta_{ij}.$$
Let $\omega_1,\dots,\omega_g$ be a normalized basis of Abelian differentials $\int_{a_i}\omega_j=\delta_{ij}$. The theta-function of the Jacobi variety
$$
 J(\Gamma)={\mathbb C}^g/\{{\mathbb Z}^g+\Omega{\mathbb Z}^g\}
$$ is given by the series
$$
 \theta(z)=\sum_{n\in{\mathbb Z}^g}\exp(\pi i<\Omega n,n>+2\pi i<n,z>),\ z\in{\mathbb C}^g,
$$
where $<n,z>=n_1z_1+\dots+n_gz_g,$ the matrix $\Omega$ has the components
$$
 \Omega_{ij}=\Omega_{ji}=\int_{b_i}\omega_j.
$$
The theta-function has the property
$$
 \theta(z+\Omega m+n)=\exp(-\pi i<\Omega m,m>-2\pi i<m,z>)\theta(z),\ m,n\in{\mathbb Z}^g.
$$
Let $\omega$ be a meromorphic one-form on $\Gamma$ with pole of order two at $q$.
We assume that $\omega$ is normalized by the condition $\int_{a_i}\omega=0$.
Let $U$ be a vector of $b$-periods of $\omega$:
$$
 U=\left(\int_{b_1}\omega,\dots,\int_{b_g}\omega\right).
$$
 Denote by $A(P):\Gamma\rightarrow J(\Gamma)$ the Abel map
$$
 A(P)=\left(\int_{P_0}^P\omega_1,\dots,\int_{P_0}^P\omega_g\right),
$$
$P_0$ is a fixed point. The function
$$
 \varphi(x,P)=\frac{\theta(A(P)-A(\gamma_1)-\dots-A(\gamma_g)-K_{\Gamma}+xU)}{\theta(A(P)-A(\gamma_1)-\dots-A(\gamma_g)-K_{\Gamma})}
 \exp(2\pi ix\int_{P_0}^P\omega),
$$
where $K_{\Gamma}$ is a vector of Riemann constants is correctly defined on $\Gamma$. The function $\varphi$ has simple poles at $\gamma_1,\dots,\gamma_g$ and
has the following form in the neighbourhood of $q$
$$
 \varphi=e^{kx}\left(\xi_0(x)+\frac{\xi_1(x)}{k}+\dots\right).
$$
After the normalization $\psi(x,P)=\frac{\varphi(x,P)}{\xi_0(x)}$ we get the Baker--Akhiezer function.

 Let us consider the following examples.

\noindent{\bf Example 1.} Let $\Gamma$ be $\bar{{\mathbb C}}={\mathbb C}P^1$, $z$ --- affine coordinate, $q=\infty$. The Baker--Akhiezer function
is
$$
 \psi(x,z)=e^{xz}.
$$
On $\Gamma\backslash\{q\}$ the function $\psi$ has no poles because the genus of $\Gamma$ equals zero.
For the meromorphic function
$$
 f(z)=z^n+c_{n-1}z^{n-1}+\dots+c_0,\ c_i\in{\mathbb C}
$$
with the pole of order $n$ at $q$ we have the operator with the constant coefficients
$$
 L(f)=\partial_x^n+c_{n-1}\partial_x^{n-1}+\dots+c_0.
$$

\noindent{\bf Example 2.}
Let $\Gamma={\mathbb C}/\{2{\mathbb Z}\omega+2{\mathbb Z}\omega'\}$ be an elliptic spectral curve and $q=0$. The Baker--Akhiezer function is
$$
 \psi=e^{-x\zeta(z)}\frac{\sigma(z+x+\gamma)}{\sigma(x+\gamma)\sigma(z+\gamma)}.
$$
We have
$$
 L_2\psi=(\partial_x^2-2\wp(x+\gamma))\psi=\wp(z)\psi,
$$
$$
 L_3\psi=\left(\partial_x^3-3\wp(x+\gamma)\partial_x-\frac{3}{2}\wp'(x+\gamma)\right)\psi=\frac{1}{2}\wp'(z)\psi,
$$
where $\sigma(z),\zeta(z),\wp(z)$ are Weierstrass functions.
Operators $L_1,L_2$ commute and satisfy the equation
$$
 L_3^2=L_2^3-\frac{g_2}{4}L_2-\frac{g_3}{4},
$$
where $g_2,g_3$ are some constants.

\noindent{\bf Example 3.} Under the degeneration $g_2,g_3\rightarrow 0$ in the Example 2 we get the caspidal spectral curve. Under this degeneration the functions
$\sigma(z),\zeta(z),\wp(z)$ become
$$
 \hat{\sigma}(z)=z,\qquad \hat{\zeta}(z)=\frac{1}{z},\qquad \hat{\wp}(z)=\frac{1}{z^2}.
$$
We get commuting differential operators with rational coefficients
$$
 \hat{\psi}(x,z)=e^{-\frac{x}{z}}\frac{z+x+\gamma}{(x+\gamma)(z+\gamma)},
$$
$$
 \hat{L}_2\hat{\psi}=\left(\partial_x^2-\frac{2}{(x+\gamma)^2}\right)\hat{\psi}=\frac{1}{z^2}\hat{\psi},
$$
$$
 \hat{L}_3\hat{\psi}=\left(\partial_x^3-\frac{3}{(x+\gamma)^2}\partial_x+\frac{3}{(x+\gamma)^3}\right)\hat{\psi}=-\frac{1}{z^3}\hat{\psi},
$$
$$
 \hat{L}_2^3=\hat{L}_3^2.
$$
\noindent{\bf Example 4.} Let us consider another degeneration of the elliptic spectral curve --- the sphere with one double point.
We identify two points on $\bar{{\mathbb C}}={\mathbb C}P^1$, $\Gamma={\mathbb C}P^1/\{a\sim -a\}$. In the case of singular spectral
curve the definition of the Baker--Akhiezer function is the same, but we should replace the genus by the arithmetic genus of the singular spectral curve.
 The arithmetic
genus of $\Gamma$ is one.
The Baker--Akhiezer function has the form

\begin{equation}\label{e3}
 \psi(x,z)=e^{xz}\left(1+\frac{\xi(x)}{z-\gamma}\right),\ q=\infty.
\end{equation}
From the identity
\begin{equation}\label{e4}
 \psi(x,a)=\psi(x,-a)
\end{equation}
we find
$$
 \xi(x)=\frac{(\gamma^2-a^2)\sinh(ax)}{a\cosh(ax)+\gamma\sinh(x)}.
$$
The functions $f(z)=z^2, g(z)=z^3-a^2z$ are rational functions on $\Gamma$ with the poles of order 2 and 3 at $q$.
Thus we have
\begin{equation}\label{e5}
 L(f)\psi=(\partial_x^2+u(x))\psi=z^2\psi,
\end{equation}
\begin{equation}\label{e6}
 L(g)\psi=\left(\partial_x^3+\left(\frac{3}{2}u(x)-a^2\right)\partial_x+\frac{3}{4}u'(x)\right)\psi=(z^3-a^2z)\psi,
\end{equation}
$$
 u(x)=\frac{2a^2(a^2-\gamma^2)}{(a\cosh(ax)+\gamma\sinh(ax))^2}.
$$
The Burchnall--Chaundy polynomial of $L_1,L_2$ is
\begin{equation}\label{e7}
 F(\lambda,\mu)=\lambda^2-\mu(\mu-a^2)^2.
\end{equation}

\noindent{\bf Example 5.}
If the spectral curve is singular, then in general $\psi$ is not a function on the spectral curve, but $\psi$ is a section
of a torsion-free sheaf on $\Gamma\backslash\{q\}$ (see \cite{Mum}). Let us consider one example. We take the same spectral curve as in the
Example 4, $\psi$ has the form (\ref{e3}), but instead of (\ref{e4}) we require
$$
 \psi(x,a)=b\psi(x,-a),\ b\in{\mathbb C}^*.
$$
In this case $\psi$ is a section of a sheaf. We have (\ref{e5})--(\ref{e7}), where
$$
 u(x)=\frac{8a^2b(a^2-\gamma^2)}{((a+ab+\gamma-b\gamma)\cosh(ax)+(a-ab+\gamma+b\gamma)\sinh(ax))^2}.
$$

\section{Method of deformation of Tyurin parameters}

Let me recall the definition of the Baker--Akhiezer function at $l>1$ (see \cite{K}). We take the {\it spectral data}
$$
 \{\Gamma,q,k^{-1},\gamma,\alpha,\omega(x)\},
$$
where $\Gamma$ is a Riemann surface of genus $g$,
$q$ is a fixed point on $\Gamma$, $k^{-1}$ is a local parameter near $q$,
$$
 \omega(x)=(\omega_1(x),\dots,\omega_{l-1}(x))
$$
is a set of smooth functions,  $\gamma=\gamma_1+\dots+\gamma_{lg}$ is a divisor on $\Gamma$, $\alpha$ is a set of vectors
$$
 \alpha_1,\dots,\alpha_{lg},\qquad
\alpha_i=(\alpha_{i,1},\dots,\alpha_{i,l-1}).
$$
The pair $(\gamma,\alpha)$ is called the {\it Tyurin parameters}. The Tyurin parameters define a stable
holomorphic vector bundle on $\Gamma$ of rank $l$ and degree
$lg$ with holomorphic sections  $\eta_1,\dots,\eta_l$. The points  $\gamma_1,\dots,\gamma_{lg}$ are the points of the linear dependence
$$
 \eta_l(\gamma_i)=\sum_{i=1}^{l-1}\alpha_{j,i}\eta_j(\gamma_i).
$$
The vector-function $\psi=(\psi_1,\dots,\psi_{l})$ is defined by the following properties.

1. In the neighbourhood of $q$ the vector-function $\psi$ has the form
$$
 \psi(x,P)=\left(\sum_{s=0}^{\infty}\xi_s(x)k^{-s}\right)\Psi_0(x,k),
$$
where $\xi_0=(1,0,\dots,0), \xi_i(x)=(\xi_i^1(x),\dots,\xi_i^l(x))$, the matrix  $\Psi_0$ satisfies the equation
$$
 \frac{d\Psi_0}{dx}=A\Psi_0,\qquad
A=\left(
\begin{array}{cccccc}
 0 & 1 & 0 & \dots  & 0 & 0\\
 0 & 0 & 1 & \dots  & 0 & 0\\
 \dots & \dots & \dots & \dots &  \dots & \dots\\
 0 & 0 & 0 & \dots  & 0 & 1 \\
 k+\omega_1 & \omega_2 & \omega_3 & \dots & \omega_{l-1} & 0
 \end{array}\right).
$$
2.  The components of $\psi$ are meromorphic functions on $\Gamma\backslash\{q\}$ with the simple poles $\gamma_1,\dots,\gamma_{lg}$,
and
$$
 {\rm Res}_{\gamma_i}\psi_j=\alpha_{i,j}{\rm Res}_{\gamma_i}\psi_{l},\quad 1\leq i\leq lg,\ 1\leq j\leq l-1.
$$
 For the rational function  $f(P)$ on $\Gamma$ with the unique pole of order $n$ at  $q$
there is a linear differential operator $L(f)$ of order $ln$ such that
$
 L(f)\psi(x,P)=f(P)\psi(x,P).
$
For two such functions $f(P),g(P)$ operators $L(f)$, $L(g)$ commute.

The main difficulty to construct operators of rank $l>1$ is the fact that the Baker--Akhiezer function is not found explicitly.
But operators can be found by the following method of deformation of Tyurin parameters.

The common eigenfunctions of commuting differential operators of rank $l$ satisfy the linear differential equation of order $l$
$$
 \psi^{(l)}(x,P)=\chi_0(x,P)\psi(x,P)+\dots+\chi_{l-1}(x,P)\psi^{(l-1)}(x,P).
$$
The coefficients $\chi_i$ are rational functions on $\Gamma$ (see \cite{K}) with the simple poles
$P_1(x),$ $\dots,$ $P_{lg}(x)\in\Gamma$, and with the following expansions in the
neighbourhood of $q$
$$
\chi_0(x,P)=k+g_0(x)+O(k^{-1}),
$$
$$
 \chi_j(x,P)=g_j(x)+O(k^{-1}), \ \
 0<j<l-1,
$$
$$
\chi_{l-1}(x,P)=O(k^{-1}).
$$
Let $k-\gamma_i(x)$ be a local parameter near $P_i(x)$. Then
$$
\chi_j=\frac{c_{i,j}(x)}{k-\gamma_i(x)}+d_{i,j}(x)+O(k-\gamma_i(x)).
$$
Functions $c_{ij}(x) ,d_{ij}(x)$ satisfy the following equations (see \cite{K}).

\vspace{0.4cm}

\begin{theorem}
\begin{equation}\label{w1}
 c_{i,l-1}(x)=-\gamma'_i(x),
\end{equation}
\begin{equation}\label{w2}
 d_{i,0}(x)=\alpha_{i,0}(x)\alpha_{i,l-2}(x)+\alpha_{i,0}(x)d_{i,l-1}(x)
 -\alpha'_{i,0}(x),
\end{equation}
\begin{equation}\label{w3}
 d_{i,j}(x)=\alpha_{i,j}(x)\alpha_{i,l-2}(x)-\alpha_{i,j-1}(x)+\alpha_{i,j}(x)d_{i,l-1}(x)
 -\alpha'_{i,j}(x), j\geq 1,
\end{equation}
where
$
 \alpha_{i,j}(x)=\frac{c_{i,j}(x)}{c_{i,l-1}(x)}, \ \ 0\leq j\leq l-1, \ 1\leq i\leq lg.
$
\end{theorem}

To find $\chi_i$ one should solve the equations (\ref{w1})--(\ref{w3}). Using $\chi_i$ one can find coefficients of the operators.
At $g=1$, $l=2$ Krichever and Novikov \cite{KN1, KN2} solved these equations and found the operators.

\begin{theorem}
The operator of order 4 has the form
$$
 L_{KN}=\left(\partial_x^2+u\right)^2+2c_x(\wp(\gamma_2)-\wp(\gamma_1))\partial_x+(c_x(\wp(\gamma_2)-\wp(\gamma_1)))_x-
 \wp(\gamma_2)-\wp(\gamma_1),
$$
where
$$
 \gamma_1(x)=\gamma_0+c(x),\ \gamma_2(x)=\gamma_0-c(x),
$$
$$
 u(x)=-\frac{1}{4c_x^2}+\frac{1}{4}\frac{c_{xx}^2}{c_x^2}+2\Phi(\gamma_1,\gamma_2)c_x-\frac{c_{xxx}}{2c_x}+
 c_x^2(\Phi_c(\gamma_0+c,\gamma_0-c)-\Phi^2(\gamma_1,\gamma_2)),
$$
$$
 \Phi(\gamma_1,\gamma_2)=\zeta(\gamma_2-\gamma_1)+\zeta(\gamma_1)-\zeta(\gamma_2),
$$
$\zeta(z),\wp(z)$ are the Weierstrass functions, $c(x)$ is an arbitrary smooth function, $\gamma_0$ is a constant.
\end{theorem}

The operator $\tilde{L}_{KN}$ commuting with $L_{KN}$ can be found from the identity
$$
 \tilde{L}_{KN}^2=4L_{KN}^3+g_2L_{KN}+g_3.
$$
The operators $L_{KN}$, $\tilde{L}_{KN}$ were studied by many authors (see \cite{GN}--\cite{Deh}).

Dixmier \cite{D} constructed an example of the commutative subalgebras in the first Weyl algebra. Let $A_1={\mathbb C}\langle p,q:[p,q]=1\rangle$  be the
Weyl algebra.

\begin{theorem} Two elements of $A_1$,
$$
 X=(p^3+q^2+h)^2+2p,
$$
$$
 Y=(p^3+q^2+h)^3+\frac{3}{2}\left(p(p^3+q^2+h)+(p^3+q^2+h)p\right),\ h\in{\mathbb C}
$$
commute and satisfy the equation
$
 Y^2=X^3-h.
$
\end{theorem}

If we substitute $p=x$, $q=-\partial_x$ in Theorem 3.3 (we can do so, because $[x,-\partial_x]=1$), then we get operators
of rank two
$$
 L_{D}=(\partial_x^2+x^3+h)^2+2x,\
$$
$$
 \tilde{L}_{D}=\left(\partial_x^2+x^3+h\right)^3+\frac{3}{2}\left(x\left(\partial_x^2+x^3+h\right)+
 \left(\partial_x^2+x^3+h\right)x\right).
$$
Operator $L_{D}$ coincides with $L_{KN}$ for some $c(x)$. Then, a natural question is how to obtain $L_D$ from $L_{KN}$ (Gelfand's problem).
 The answer is given in the following theorem by Grinevich \cite{G}.

\begin{theorem}
Operator $L_{KN}$ corresponding to the curve $w^2=4z^3+g_2z+g_3$ has rational coefficients if and only if
$$
 c(x)=\int_{q(x)}^{\infty}\frac{dt}{\sqrt{4t^3+g_2t+g_3}},
$$
where $q(x)$ is a rational function. If $\gamma_0=0$ and $q(x)=x$, then the operator $L_{KN}$ coincides with $L_D$.
\end{theorem}

Grinevich and Novikov \cite{GN} found conditions when $L_{KN}$ is self-adjoint.

\begin{theorem}
Operator $L_{KN}$ is self-adjoint if and only if $\wp(\gamma_1)=\wp(\gamma_2)$.
\end{theorem}

Spectral properties of operators with periodic coefficients of rank $l>1$ with a spectral curve of arbitrary genus were studied by Novikov in \cite{N}.

\section{Self-adjoint operators of rank two}
If $l=2$, then
\begin{equation}\label{u1}
 \psi''=\chi_0\psi+\chi_1\psi'.
\end{equation}
In the neighbourhood of $q$ we have the expansions
\begin{equation}\label{u3}
\chi_0=\frac{1}{k}+a_0(x)+a_1(x)k+O(k^2),\quad \chi_1=b_1(x)k+b_2(x)k^2+O(k^3).
\end{equation}
Functions $\chi_0,\chi_1$ have $2g$ simple poles $P_1(x),\dots,P_{2g}(x)$, and by Theorem 3.1

\begin{equation}\label{u4}
\chi_0(x,P)=\frac{-\alpha_{i,0}(x)\gamma'_i(x)}{k-\gamma_i(x)}+d_{i,0}(x)+O(k-\gamma_i(x)),
\end{equation}
\begin{equation}\label{u5}
\chi_1(x,P)=\frac{-\gamma'_i(x)}{k-\gamma_i(x)}+d_{i,1}(x)+O(k-\gamma_i(x)),
\end{equation}
\begin{equation}\label{u6}
d_{i,0}(x)=\alpha_{i,0}^2(x)+\alpha_{i,0}(x)d_{i,1}(x)-\alpha'_{i,0}(x).
\end{equation}
Let $\Gamma$ be the hyperelliptic spectral curve
\begin{equation}\label{v1}
 w^2=F_g(z)=z^{2g+1}+c_{2g}z^{2g}+\dots+c_0,
\end{equation}
and $q=\infty\in \Gamma,$ $k=\frac{1}{\sqrt{z}}$. The curve $\Gamma$ has the involution
$$
\sigma:\Gamma\rightarrow\Gamma,\quad  \sigma(z,w)=(z,-w).
$$
Let us find coefficients of the operator of order 4 corresponding to  $z$
$$
 L_4\psi=(\partial_x^4+f_2(x)\partial_x^2+f_1(x)\partial_x+f_0(x))\psi=z\psi.
$$
From (\ref{u1}) it follows that the fourth derivative of $\psi$ is
$$
 \psi^{(4)}=(\chi_0^2+\chi_1\chi_0'+\chi_0(\chi_1^2+2\chi_1')+\chi_0'')\psi+
 (\chi_1^3+2\chi_0'+\chi_1(2\chi_0+3\chi_1')+\chi_1'')\psi'.
$$
With the help of (\ref{u1}) and the last identity we rewrite $L_4\psi=z\psi$ in the form
$
P_1\psi+P_2\psi'=z\psi,
$
where
$$
 P_1=f_0+f_2\chi_0+\chi_0^2+\chi_1\chi_0'+\chi_0(\chi_1^2+2\chi_1')+\chi_0'',
$$
$$
 P_2=f_1+f_2\chi_1+\chi_1^3+2\chi_0'+\chi_1(2\chi_0+3\chi_1')+\chi_1''.
$$
It gives
\begin{equation}\label{e11}
 P_1=z=\frac{1}{k^2},\qquad \ P_2=0.
\end{equation}
From (\ref{u3}) we have
$$
 P_1-\frac{1}{k^2}=\frac{f_2+2a_0}{k}+(f_0+a_0(f_2+a_0)+2(a_1+b_1')+a_0'')+O(k)=0,
$$
$$
 P_2=(f_1+2(b_1+a_0'))+O(k)=0.
$$
Hence $f_0=a_0^2-2a_1-2b_1'-a_0'',\ f_1=-2(b_1+a_0'),\ f_2=-2a_0.$
If $b_1=0$, then the operator $L_4$ is self-adjoint
$$
 L_4=(\partial_x^2+V(x))^2+W(x),
$$
where $V(x)=-a_0(x),$ $W=-2a_1(x)$.
If $\chi_1(x,P)=\chi_1(x,\sigma(P)),$ then
$$\chi_1=\sum_{s>1}b_{2s}k^{2s},$$
hence, $L_4$ is self-adjoint (the inverse is also true, see Theorem 4.1).
Assume that $\chi_1$ is invariant under $\sigma$, then by (\ref{u3})--(\ref{u5}) we have
$$
 \chi_0=-\frac{H_1(x)\gamma'_1(x)}{z-\gamma_1(x)}-\dots-
 \frac{H_g(x)\gamma'_g(x)}{z-\gamma_g(x)}+\frac{w(z)}{(z-\gamma_1(x))\dots(z-\gamma_g(x))}+\kappa(x),
$$
$$
 \chi_1(x,P)=-\frac{\gamma'_1(x)}{z-\gamma_1(x)}-\dots-
 \frac{\gamma'_g(x)}{z-\gamma_g(x)},
$$
where $H_i(x),\kappa(x)$ are some functions. In the neighbourhood of $q$ the function $\chi_0$ has the expansion
$$
\chi_0=\frac{1}{k}+\kappa+\left(\gamma_1+\dots+\gamma_g+\frac{c_{2g}}{2}\right)k+O(k^2).
$$
Hence,
\begin{equation}\label{u8}
 V=-\kappa,\qquad W=-2(\gamma_1+\dots+\gamma_g)-c_{2g}, \ 1\leq i\leq g.
\end{equation}
Functions $\chi_0$, $\chi_1$ have simple poles at $P_i^{\pm}=(\gamma_i,\pm\sqrt{F_g(\gamma_i)})$.
Denote by $\alpha_{i,0}^{\pm}(x)$, $d_{i,0}(x)^{\pm},$ $d_{i,1}(x)^{\pm}$ coefficients of expansions of $\chi_0$, $\chi_1$ at $P_i^{\pm}$. From
(\ref{u6}) we have
$$
 l_i^{\pm}=d_{i,0}^{\pm}-((\alpha_{i,0}^{\pm})^2+\alpha_{i,0}^{\pm}d_{i,1}^{\pm}-(\alpha_{i,0}^{\pm})')=0.
$$
From $l_i^{+}-l_i^-=0$ one can express $H_i$ through $\gamma_1,\dots,\gamma_g$ and its derivatives. From $l_i^{+}+l_i^-=0$
one can express $\kappa(x)$ through $\gamma_i,H_i,\ i=1,\dots,g$ and its derivatives. Thus, we can reduce the system (\ref{u6})
to the system of $g-1$ equations on $\gamma_i$.

Let us consider two examples at $g=1, 2$.

\subsection{Dixmier operators} {}

Let $\Gamma$ be the elliptic curve
$$
 w^2=F_1(z)=z^3+c_2z^2+c_1z+c_0.
$$
We have
$$
 \chi_0=-\frac{H_1(x)\gamma_1'(x)}{z-\gamma_1(x)}
 +\frac{w(z)}{z-\gamma_1(x)}+\kappa(x),\qquad
 \chi_1(x,P)=-\frac{\gamma_1'(x)}{z-\gamma_1(x)}.
$$
The coefficients of expansions of $\chi_0,\chi_1$ at $P_1^{\pm}$ are
$$
 \alpha_{1,0}^{\pm}=H_1(x)\mp\frac{w(\gamma_1)}{\gamma_1'},\qquad
  d_{1,0}^{\pm}=\kappa(x)\pm\frac{F_1'(\gamma_1)}{2w(\gamma_1)},\qquad
  d_{1,1}^{\pm}=0,
$$
hence,
$$
 l^{\pm}=\kappa(x)-\left(H_1(x)\mp\frac{w(\gamma_1)}{\gamma_1'}\right)^2+H_1'(x)\pm\frac{w(\gamma_1)\gamma_1''}{(\gamma_1')^2}=0.
$$
From $l^+-l^-=0$ we find
$
 H_1(x)=-\frac{\gamma_1''(x)}{2\gamma_1'(x)}.
$

\noindent From $l^++l^-=0$ we get
$
 \kappa(x)=\frac{4F_1(\gamma_1)-(\gamma_1'')^2+2\gamma_1'\gamma_1'''}{4(\gamma_1')^2}.
$

 At
$$\gamma_1=-h_3x-h_2,\ F_1=z^3+2 h_2z^2+z(h_2^2+h_1h_3)+h_3(h_1h_2-h_0h_3)$$
we have
$$
 \chi_0=\frac{\sqrt{F_1(z)}}{z+h_3x+h_2}-(h_3x^3+h_2x^2+h_1x+h_0),\qquad \chi_1=\frac{h_3}{z+h_3x+h_2},
$$
$$
 V=-\kappa(x)=h_3x^3+h_2x^2+h_1x+h_0,\qquad W=2h_3x.
$$
Thus we get the operator
$$
 L^{^{\sharp}}_4=(\partial_x^2+h_3x^3+h_2x^2+h_1x+h_0)^2+2h_3x.
$$
At $g=1, h_0=h_1=h_2=0, h_3=1$. The operator $L^{^{\sharp}}_4$ coincides with the Dixmier operator.

\subsection{Spectral curves of genus two}

\vspace{0.4cm}

Let $\Gamma$ be a spectral curve of genus two
$$
 w^2=F_2(z)=z^5+c_4z^4+c_3z^3+c_2z^2+c_1z+c_0.
$$
We have
$$
 \chi_0=-\frac{H_1(x)\gamma'_1(x)}{z-\gamma_1(x)}-\frac{H_2(x)\gamma'_2(x)}{z-\gamma_2(x)}
 +\frac{w(z)}{(z-\gamma_1(x))(z-\gamma_2(x))}+\kappa(x),
$$
$$
 \chi_1(x,P)=-\frac{\gamma'_1(x)}{z-\gamma_1(x)}-\frac{\gamma'_2(x)}{z-\gamma_2(x)}.
$$
The coefficients of expansions of $\chi_0$, $\chi_1$ at $P_1^{\pm}(x), P_2^{\pm}(x)$ are
$$
 \alpha_{1,0}^{\pm}=H_1(x)\mp\frac{w(\gamma_1)}{(\gamma_1-\gamma_2)\gamma_1'},\ \
 \alpha_{2,0}^{\pm}=H_2(x)\pm\frac{w(\gamma_2)}{(\gamma_1-\gamma_2)\gamma_2'},
$$
$$
 d_{1,0}^{\pm}=\mp\frac{w(\gamma_1)}{(\gamma_1-\gamma_2)^2}+\kappa(x)\pm\frac{F_2'(\gamma_1)}{2w(\gamma_1)(\gamma_1-\gamma_2)}-\frac{H_2\gamma_2'}{\gamma_1-\gamma_2},
$$
$$
 d_{2,0}^{\pm}=\mp\frac{w(\gamma_2)}{(\gamma_1-\gamma_2)^2}+\kappa(x)\pm\frac{F_2'(\gamma_2)}{2w(\gamma_2)(\gamma_2-\gamma_1)}+\frac{H_1\gamma_1'}{\gamma_1-\gamma_2},
$$
$$
 d_{1,1}^{\pm}=-\frac{\gamma_2'(x)}{\gamma_1(x)-\gamma_2(x)},\ d_{2,1}^{\pm}=\frac{\gamma_1'(x)}{\gamma_1(x)-\gamma_2(x)}.
$$
Equations $l_1^{\pm}=0,l_2^{\pm}=0$ have the form
$$
 l_1^{\pm}=-\frac{1}{(\gamma_1-\gamma_2)^2(\gamma_1')^2}
 \left(F_2(\gamma_1)+(\gamma_1-\gamma_2)(\gamma_1')^2\left((H_2-H_1)\gamma_2'\right.\right.
$$
$$
 \left.\left.+(H_1^2-\kappa-H_1')(\gamma_1-\gamma_2)\right)+w(\gamma_1)(\pm 2\gamma_1'\gamma_2'\pm(2H_1\gamma_1'+\gamma_1'')(\gamma_2-\gamma_1))\right)=0,
$$
$$
 l_2^{\pm}=-\frac{1}{(\gamma_1-\gamma_2)^2(\gamma_2')^2}
 \left(F_2(\gamma_2)+(\gamma_1-\gamma_2)(\gamma_2')^2((H_2-H_1)\gamma_1'\right.+
$$
$$
 \left.(H_2^2-\kappa-H_2')(\gamma_1-
 \gamma_2))+w(\gamma_2)(\pm2\gamma_1'\gamma_2'\pm(2H_2\gamma_2'+\gamma_2'')(\gamma_1-\gamma_2))\right)=0,
$$
From the equations $l_1^{+}-l_1^{-}=0$ and $l_2^{+}-l_2^{-}=0$ we find
$$
 H_1(x)=\frac{\gamma_2'}{\gamma_1-\gamma_2}-\frac{\gamma_1''}{2\gamma_1'},
\qquad
 H_2(x)=-\frac{\gamma_1'}{\gamma_1-\gamma_2}-\frac{\gamma_2''}{2\gamma_2'}.
$$
From the equations $l_1^{+}+l_1^{-}=0$ and $l_2^{+}+l_2^{-}=0$ we can find $\kappa(x)$ by two ways
$$
 \kappa(x)=
 \frac{F_2(\gamma_1)+(\gamma_1-\gamma_2)(\gamma_1')^2((H_2-H_1)\gamma_2'+\gamma_1(H_1^2-H_1')-\gamma_2(H_1^2-H_1'))}{(\gamma_1-\gamma_2)^2\gamma_1'^2},
$$
$$
 \kappa(x)=
 \frac{F_2(\gamma_2)+(\gamma_2-\gamma_1)(\gamma_2')^2((H_1-H_2)\gamma_1'+\gamma_2(H_2^2-H_2')-\gamma_1(H_2^2-H_2'))}{(\gamma_2-\gamma_1)^2\gamma_2'^2}.
$$
We get the equation on $\gamma_1$ and $\gamma_2$ (see \cite{M3})
\begin{equation}\label{ur}
 4((\gamma_1')^2F_2(\gamma_2)-(\gamma_2')^2F_2(\gamma_1))-4(\gamma_1')^4(\gamma_2')^2+(\gamma_1-\gamma_2)^2(\gamma_2')^2(\gamma_1'')^2
\end{equation}
$$
 +2(\gamma_1-\gamma_2)(\gamma_1')^3\gamma_2'\gamma_2''+2(\gamma_1-\gamma_2)\gamma_1'(\gamma_2')^2(\gamma_2'\gamma_1''+(\gamma_2-\gamma_1)\gamma_1''')+
$$
$$
 (\gamma_1')^2(4(\gamma_2')^4+6(\gamma_1-\gamma_2)(\gamma_2')^2(\gamma_1''+\gamma_2'')+
 (\gamma_1-\gamma_2)^2(2\gamma_2'\gamma_2'''-(\gamma_2'')^2))=0.
$$
O.I. Mokhov also considered the case of self-adjoint operators of rank two, corresponding to a curve of genus two. In particular,
he also reduced the equations on Tyurin parameters to one equation on two functions (see introduction in \cite{Mokh}).

Equation (\ref{ur})  has the following solution
$$
 \gamma_1=\frac{1}{2}(-5h_2-3h_3x+p),\qquad
 \gamma_2=\frac{1}{2}(-5h_2-3h_3x-p),\
$$
$$
 p=3\sqrt{h_2^2-2h_3(2h_1-h_2x)-3h_3^2x^2},
$$
$$
  F_2=z^5+10h_2z^4+(33h_2^2+21h_1h_3) z^3+(40h_2^3+117h_1h_2h_3+27h_0h_3^3)z^2+
$$
$$
 4(4h_2^4+36h_1h_2^2h_3+27h_3^2(h_1^2+h_0h_2))z+
 3h_3(36h_1^2h_2h_3+27h_3^3+4h_1(4h_2^3+27h_0h_3^2)).
$$
Formula (\ref{u8}) gives us
$
 V=-\kappa=h_3x^3+h_2x^2+h_1x+h_0,\ W=-2(\gamma_1+\gamma_2)-10h_2=6h_3x.
$
Thus we get the operator
$$
 L^{^{\sharp}}_4=(\partial_x^2+h_3x^3+h_2x^2+h_1x+h_0)^2+6h_3x.
$$
V.V. Sokolov noticed that the equation (\ref{ur}) has also polynomial solutions of degree two.

\subsection{Hyperelliptic spectral curves}
As we showed above, if
$\chi_1$ is invariant under the involution $\sigma$, then the operator $L_4$ is
self-adjoint. S.P. Novikov has proposed the conjecture that the inverse is also true.

\begin{theorem}[M., \cite{M4}] The operator $L_4$ is self-adjoint if and only if
\begin{equation}\label{u2}
 \chi_1(x,P)=\chi_1(x,\sigma(P)).
\end{equation}
\end{theorem}

At $g=1$ Theorem 4.1 is equivalent to the Theorem 3.5 by Grinevich and Novikov.

Let us assume that the operator $L_4$ is self-adjoint
$
 L_4=(\partial_x^2+V(x))^2+W(x),
$
then the functions $\chi_0,\chi_1$ have simple poles at some points
$$\left(\gamma_i(x),\pm\sqrt{F_g(\gamma_i(x))}\right),\ 1\leq i\leq g.$$

\begin{theorem}[M., \cite{M4}]
If operator $L_4$ is self-adjoint, then
$$
 \chi_0=-\frac{1}{2}\frac{Q''}{Q}+\frac{w}{Q}-V, \qquad \chi_1=\frac{Q'}{Q},
$$
where $Q=(z-\gamma_1(x))\dots(z-\gamma_g(x))$. Functions $Q,V,W$ satisfy the equation
\begin{equation}\label{e1}
 4F_g(z)=4(z-W)Q^2-4V(Q')^2+(Q'')^2-2Q'Q^{(3)}
 +2Q(2V'Q'+4VQ''+Q^{(4)}),
\end{equation}
where $Q',Q'',Q^{(k)}$ mean $\partial_xQ,\partial_x^2Q,\partial_x^kQ.$
\end{theorem}

To find self-adjoint operators $L_4, L_{4g+2}$ it is enough to solve the equation (\ref{e1}).

\vspace{0.4cm}

\noindent{\bf Corollary 1} {} {\it The functions $Q,W,V$ satisfy the equation
$$
 Q^{(5)}+4VQ^{3}+2Q'(2z-2W-V'')+6V'Q''-2QW'=0.
$$}
Let us substitute $z=\gamma_j$ in (\ref{e1}). This gives
$$
 V(x)=\left(\frac{(Q'')^2-2Q'Q^{(3)}-4F_g(z)}{4(Q')^2}\right)\mid_{z=\gamma_j}.
$$
We get $g-1$ equations on $\gamma_1(x),\dots,\gamma_g(x)$.

\vspace{0.4cm}

\noindent{\bf Corollary 2} {} {\it The functions $\gamma_1(x),\dots,\gamma_g(x)$ satisfy the equations
\begin{equation}\label{e2}
 \left(\frac{(Q'')^2-2Q'Q^{(3)}-4F_g(z)}{4(Q')^2}\right)\mid_{z=\gamma_j}=\left(\frac{(Q'')^2-2Q'Q^{(3)}-4F_g(z)}{4(Q')^2}\right)\mid_{z=\gamma_k}.
\end{equation}}
At $g=2$ the equation (\ref{e2}) coincides with the equation (\ref{ur}).
In \cite{M4} partial solutions of the equation (\ref{e1}) are found for arbitrary $g$.
 These solutions give the first examples
of commuting differential operators of rank greater than one corresponding to a spectral curve of arbitrary genus.

\begin{theorem}[M., \cite{M4}]
The operator
$$
 L^{^{\sharp}}_4=(\partial_x^2+h_3x^3+h_2 x^2+h_1x+h_0)^2+g(g+1)h_3x, \qquad h_3\ne 0
$$
commutes with a differential operator
$L_{4g+2}^{^{\sharp}}$ of order $4g+2$. The operators
$L_4^{^{\sharp}},$ $L_{4g+2}^{^{\sharp}}$ are operators of rank two.
For generic values of parameters $(h_0,h_1,h_2,h_3)$ the spectral
curve is a nonsingular hyperelliptic curve of genus $g$.
\end{theorem}

 Operators $L^{^{\sharp}}_4,L_{4g+2}^{^{\sharp}}$ define
commutative subalgebras in the first Weyl algebra $A_1$.

\section{Concluding remarks and questions}

1. G. Latham and E. Previato \cite{LP} proved the following statement. Let $L_4$ and $L_6$ be operators of rank two corresponding
to an elliptic spectral curve. Then there are $z_0,w_0\in{\mathbb C}$ such that
$$
 L_4-z_0=A_2T,\qquad L_6-w_0=A_4T
$$
for some operators $A_2,A_4,T$ and such that
$$
 \tilde{L}_4=TA_2=T(L_4-z_0)T^{-1},\qquad \tilde{L}_6=TA_4=T(L_6-w_0)T^{-1}
$$
are commuting self-adjoint operators of rank two. In other words, in the case of elliptic spectral curves all operators of rank two up to the conjugation
by  operators of the second order
are self-adjoint operators. In the beginning of the 1980's O.I. Mokhov proved a similar result where $T$ is an operator of the first order (this result
is not published).
It would be very interesting to check this property for operators $L_4$, $L_{4g+2}$ corresponding to an hyperelliptic spectral curve
of genus $g$.

2. The group of automorphisms of the first Weyl algebra $Aut(A_1)$ acts on the moduli spaces of operators with polynomial coefficients.
 For example, with the help of the automorphism
$$
 \varphi_1(x)=\alpha x+\beta\partial_x,\qquad \varphi_1(\partial_x)=\gamma x+\delta\partial_x,\qquad
 \left(
 \begin{array}{cc}
 \alpha & \beta \\
 \gamma & \delta
 \end{array}\right)\in {\rm SL}_2
$$
one can get from $L_4^{^{\sharp}},L^{^{\sharp}}_{4g+2}$ the operators of rank 3. Another example of automorphisms are
$$
 \varphi_2(x)=x+P_1(\partial_x),\qquad \varphi_2(\partial_x)=\partial_x,
$$
$$
 \varphi_3(x)=x,\qquad \varphi_3(\partial_x)=\partial_x+P_2(x),
$$
where $P_1,P_2$ are polynomials. Dixmier \cite{D} proved that $ Aut (A_1)$ is generated by $\varphi_i$. It would be very
interesting to understand how $ Aut(A_1)$ acts on the spectral data.

3. Theorem 4.3 means that the equation
$
 Y^2=X^{2g+1}+c_{2g}X^{2g}+\dots+c_0
$
has nonconstant solutions $X,Y\in A_1$ for some $c_i$.

It is easy to see that the group $ Aut(A_1)$ preserves the space of
all such solutions, i.e. if $(X,Y)$ is a solution to the
polynomial equation above, with $X,Y \in A_1 $, then
$ (\varphi(X), \varphi(Y))$ is also a solution for any $ \varphi \in
Aut(A_1)$. Then, a natural question is to describe the orbits of $ Aut(A_1) $
in the space of solutions under the action of $ Aut(A_1) $.

Yu. Berest has proposed the following conjecture: If $ g > 1 $, then
there are only finitely many such orbits, i.e. the equation
$
f(X,Y)=\sum_{i,j=0}^k\alpha_{ij}X^iY^j=0
$
with generic $ \alpha_{ij} \in {\mathbb C} $ has at most finitely many
solutions  in $ A_1 $ up to the action of $ Aut(A_1) $.

4. Let me recall the Dixmier conjecture:
$
  End (A_1)= Aut (A_1).
$
If one describe all orbits of $ Aut(A_1) $ in the space of solutions for the equation $f(X,Y)=0$, then this gives a chance to compare $End (A_1)$ and $Aut (A_1)$.
For example, if there is only one orbit, then $ End (A_1)= Aut (A_1).$
For this reason it is important to find all solutions $X,Y\in A_1$ for one concrete equation and to study the action of $ Aut (A_1)$.
For example, one can take the simplest equation
$
 Y^2=X^3+1.
$

\end{document}